\documentclass[reqno,12pt]{article}
\usepackage{amsmath,amsfonts,amssymb,amstext,amscd,eucal}
\usepackage[all]{xy}
\usepackage{mathtools}
\usepackage{slashed}
\usepackage{float}
\usepackage{simplewick}
\usepackage{xcolor,colortbl}

\newcommand{\mathsym}[1]{{}} 
\usepackage[colorlinks=true,
            linkcolor=blue,
            urlcolor=blue,
            citecolor=blue]{hyperref}
\usepackage{enumerate}

\usepackage{graphicx}
\usepackage{bm}
\usepackage{cite,hyperref}

\DeclareMathAlphabet{\pazocal}{OMS}{zplm}{m}{n}
\usepackage[active]{srcltx}
\makeatletter \@addtoreset{equation}{section}

\makeatletter\renewcommand\section{\@startsection {section}{1}{\z@}%
                                   {-3.5ex \@plus -1ex \@minus -.2ex}
                                   {2.3ex \@plus.2ex}%
                                   {\normalfont\large\bfseries}}
\renewcommand\subsection{\@startsection{subsection}{2}{\z@}%
                                     {-3.25ex\@plus -1ex \@minus -.2ex}%
                                     {1.5ex \@plus .2ex}%
                                     {\normalfont\bfseries}}

\newcommand{\be}{\begin{equation}}
\newcommand{\ee}{\end{equation}}
\newcommand{\bea}{\begin{eqnarray}}
\newcommand{\eea}{\end{eqnarray}}
\newcommand{\bse}{\begin{subequations}}
\newcommand{\ese}{\end{subequations}}
\newcommand{\beqa}{\begin{eqnarray}}
\newcommand{\eeqa}{\end{eqnarray}}
\newcommand{\beqar}{\begin{eqnarray*}}
\newcommand{\eeqar}{\end{eqnarray*}}
\newcommand{\bi}{\begin{itemize}}
\newcommand{\ei}{\end{itemize}}
\newcommand{\bn}{\begin{enumerate}}
\newcommand{\en}{\end{enumerate}}

\newcommand{\ba}{\begin{array}}
\newcommand{\ea}{\end{array}}
\newcommand{\bc}{\begin{center}}
\newcommand{\ec}{\end{center}}

\definecolor{darkgreen}{rgb}{0,0.3,0}
\definecolor{darkblue}{rgb}{0,0,0.3}
\definecolor{darkred}{rgb}{0.7,0,0}
\definecolor{VioletRed4}{rgb}{0.55,0.13,0.32}
\definecolor{VioletRed}{rgb}{0.82,0.13,0.56}
\definecolor{VioletRed2}{rgb}{0.93,0.23,0.55}
\definecolor{DeepPink}{rgb}{1.00,0.08,0.58}
\definecolor{DeepPink2}{rgb}{0.93,0.07,0.54}

\begin{document}
\setcounter{footnote}{0}
\renewcommand{\baselinestretch}{1.05}
\newcommand{\email}[1]{\footnote{\href{mailto:#1}{#1}}}

\title{\bf\Large{Impossibility of obtaining a CP-violating Euler-Heisenberg effective theory from a viable modification of QED}}

\author{\bf{M.~Ghasemkhani}\email{ghasemkhani@ipm.ir} $^{a}$, \bf{V.~Rahmanpour}\email{v.rahmanpour@mail.sbu.ac.ir} $^{a}$, R.~Bufalo\email{rodrigo.bufalo@ufla.br} $^{b}$, and A.~Soto\email{arsoto1@uc.cl} $^{c}$ \\\\
\textit{\small$^a$ Department of Physics, Shahid Beheshti University, 1983969411, Tehran, Iran}\\
\textit{\small$^b$ Departamento de F\'isica, Universidade Federal de Lavras,}\\
\textit{ \small Caixa Postal 3037, 37200-900 Lavras, MG, Brazil}\\
\textit{\small$^c$ School of Mathematics, Statistics and Physics, Newcastle University,}\\
 \textit{ \small Newcastle upon Tyne, NE1 7RU, UK}}
\maketitle

\begin{abstract}
In this paper, we examine the CP-violating term of the Euler-Heisenberg action.
We focus in the aspects related with the generation of such a term from a QED-like model in terms of the effective action approach.
In particular, we show that the generation of the CP-violating term is closely related with both of vector and axial fermionic bilinears.
Although, these anomalous models are not a viable extension of QED, we argue that the CP-violating term in the photon sector is obtained only from this class of models, and not from any fundamental field theory.
\end{abstract}

\newpage
\section{Introduction}

In recent years we have seen an increasing interest in the (quantum) nonlinear extension of the Maxwell's electromagnetic theory, mainly focused in the light-by-light scattering.
Almost 80 years have passed since the early proposals of the light-by-light scattering in QED  \cite{heisenberg,euler-kockel,heisenberg-euler,karplus-neuman}, until the conceptual proposal of light-by-light scattering in ultraperipheral heavy-ion collisions  \cite{dEnterria:2013zqi} and its experimental verification by ATLAS Collaboration \cite{ATLAS:2017fur,ATLAS:2019azn}.
Other interests in these nonlinear corrections worth mentioning are low-energy experiments, such as PVLAS \cite{Ejlli:2020yhk} and BMV \cite{Cadene:2013bva}, built to detect the presence of vacuum magnetic birefringence.

The best framework to derive these nonlinear corrections is in the context of (quantum) effective field theories , resulting in the phenomena of (quantum) self-coupling of electromagnetic waves in the vacuum \cite{Schwinger:1951nm,Dittrich:2000zu,Dunne:2004nc,Preucil:2017,Quevillon:2018mfl}.
One can obtain the usual Euler-Heisenberg action from QED in the presence of an external gauge field \cite{Schwinger:1951nm}
\begin{align}
\pazocal{L}_{\rm eff} = &- 4\pazocal{F} + \frac{1}{8\pi ^2} \int _0^\infty \frac{ds}{s^3} e^{-ism^2} \Bigg\{ \frac{8e^2s^2}{3} \pazocal{F} -1 \cr
&+ 4 (es)^2 |\pazocal{G}| \cot \left[2es \left(\sqrt{\pazocal{F}^2+\pazocal{G}^2} + \pazocal{F} \right)^{\frac{1}{2}}\right] \coth \left[2es \left(\sqrt{\pazocal{F}^2+\pazocal{G}^2} + \pazocal{F} \right)^{\frac{1}{2}}\right] \Bigg\},
\end{align}
which, in the weak-field regime, can be cast as
\begin{align}\label{eq_eh}
\pazocal{L}_{\rm eff} = \frac{\alpha^2}{90m_e^4} \pazocal{F}^{2} +  \frac{7\alpha^2}{360m_e^4} \pazocal{G}^{2},
\end{align}
where we have the following gauge invariant quantities
\begin{align}
\pazocal{F}&= F_{\mu\nu}F^{\mu\nu}=-2(\mathbf{E^2-B^2}),\nonumber\\
\pazocal{G}&=  G_{\mu\nu}F^{\mu\nu}=4(\mathbf{E.B}),
\label{eq:b11}
\end{align}
and $G_{\mu\nu}=\frac{1}{2} \varepsilon_{\mu\nu\rho\sigma}F^{\rho\sigma}$ is the dual of the field strength tensor $F_{\mu\nu}=\partial_{\mu}A_{\nu}-\partial_{\nu}A_{\mu}$.

This same expression \eqref{eq_eh} can also be found from perturbative analysis when evaluating the box diagram with fermionic internal lines (similar expression is found for bosonic matter, but with different numerical coefficients \cite{Dittrich:2000zu,Dunne:2004nc,Preucil:2017}).

Since the experimental endeavour in recent years led to a great progress in the understanding of the non linear regime of the electromagnetic field, one can naturally ask about possible extensions of the Euler-Heisenberg action.
In general, these generalizations may be intrinsically related with the breaking of some symmetry. In particular, we shall focus our discussion on discrete symmetries.
In light of this reasoning, we can draw some considerations: if the action is invariant by $C$, $P$, and $T$ transformations, the $\alpha^2$-order nonlinear corrections are described by \eqref{eq_eh}.
However, if we allow CP violation, the term $\pazocal{F} \pazocal{G}$ must be also added.
The (quantum) generation of this CP-violating sector is precisely the point of our interest.
The CP-violating term $\pazocal{F} \pazocal{G}$ in nonlinear models  has been scrutinized in the phenomenological analysis involving the measurement of the vacuum magnetic birefringence  \cite{Ejlli:2020yhk,Cadene:2013bva}, as well as some other aspects \cite{Millo:2008ug,Gorghetto:2021luj}.
Hence, its generation in the photon effective action deserves further investigation, because it poses an important aspect of the photon's quantum dynamics.
With this aim, we start Sec.~\ref{sec2} by establishing the main aspects of the model in the context of effective field theory.
In particular, we discuss how the  generation of this CP-violating term $\pazocal{F} \pazocal{G}$ through the one-loop quantum corrections is related with the presence of an axial coupling among the gauge field and the matter field.
Actually, it is the mixture of axial and vector photon-matter couplings responsible to engender the CP-violating effects in the nonlinear action.
Finally, we present our conclusions and final remarks in Sec.~\ref{conc}.
\section{Effective theory: the model and main features}
\label{sec2}
It is well known that the Lagrangian density of QED
\begin{equation}
{\pazocal{L}}_{\psi}= \bar{\psi}\gamma^{\mu}\big(i\partial_{\mu}-e A_{\mu}\big)\psi-m\bar{\psi}\psi,
\label{eq:a1}
\end{equation}
is responsible to generate the $C$, $P$, and $T$ invariant Euler-Heisenberg action
\eqref{eq_eh}.

On the other hand, the CP-violating effects in the four photon interactions do not exist in the Standard Model at tree-level.
But the Standard Model contains  sources of photon interactions via CP-violation in terms of multi-loop level from the weak interactions (CP-violating phase of the CKM matrix), or by the tiny strong CP phase, and in both cases they are negligibly small \cite{Millo:2008ug,Gorghetto:2021luj}.

Hence, in order to generate perturbatively the CP-violating term $\pazocal{F}\pazocal{G}$ in the photon sector it is necessary that, at least, one of these discrete symmetries is broken.
This can be achieved by adding new couplings related with physics beyond the standard model.
In particular, as we will discuss, these couplings are necessarily axial and therefore they break both C and P symmetries in the photon-matter couplings automatically.
A main consequence of the C-violation is that the Furry's theorem is not satisfied by this model, as we expect.
As a matter of fact, only fermionic bilinear covariants such as axial-vector, axial-tensor, etc, coupling with the photon can generate the CP-violating term $\pazocal{F}\pazocal{G}$ in the photon sector, see \cite{Gorghetto:2021luj} for further details.
Hence, we observe that only anomalous fermionic models are related with the CP-violating phase of the photon sector.
We shall examine these aspects below.

The simplest interacting coupling that we can consider in order to induce $\pazocal{F}\pazocal{G}$ term is adding the axial-vector one \cite{Yamashita:2017}
\begin{equation}
{\pazocal{L}}_{int}=-\bar\psi\gamma^{\mu}(g_{v}+g_{a}\gamma^{5})A_{\mu}\psi=-\bar\psi\gamma^{\mu}(\beta e^{\alpha\gamma^{5}})A_{\mu}\psi,
\label{eq:a2}
\end{equation}
where $g_{v}$ and $g_{a}$ refer to the coupling of the external gauge field to the vector and axial vector current, respectively.
We have also introduced the following parametrization $g_{v}+g_{a}\gamma^{5}=\beta e^{\alpha\gamma^{5}}$ which allows a clear visualization of the contribution of the parity-conserving and parity-violating (axial) terms in the perturbative analysis.
Hence, the Feynman rule for the fermion-photon interaction is simply given by $-i\beta\gamma^{\mu}e^{\alpha\gamma^{5}}$, having the same structure as the usual QED with the additional factor $e^{\alpha\gamma^{5}}$.
Moreover, we observe that the axial part of the Lagrangian \eqref{eq:a2} violates the parity (P) and charge conjugation (C) symmetry, which is odd under P and C.
Hence, unlike the usual QED, this model does not respect the parity and charge conjugation symmetry.
One can observe from \eqref{eq:a2}
that the electromagnetic coupling is no longer $e$.

In ref.\cite{Ghasemkhani:2021kzf}, we have explicitly shown how the couplings in \eqref{eq:a2}, by evaluating the box diagram, induce a CP-violating term $\pazocal{F}\pazocal{G}$ in the Euler-Heisenberg effective action.

Another possibility to generate $\pazocal{F}\pazocal{G}$ is to consider the dipole operators  \cite{Gorghetto:2021luj}
\begin{equation} \label{eq_dip}
{\pazocal{L}}_{\rm int}  = g \bar{\psi}\sigma_{\mu\nu} \psi F^{\mu\nu} +  ig' \bar{\psi}\sigma_{\mu\nu} \gamma_5 \psi F^{\mu\nu},
\end{equation}
where $g$ and $g'$ are the fermion magnetic and electric dipole moments, respectively.
While electron's magnetic moment (spin) is a fundamental property, its electric dipole can arise only from quantum corrections; actually, one has a very small value from the CP-violating components of the CKM matrix in the standard model \cite{Khriplovich:1997ga}).
Hence, although the presence of the electron's electric dipole in QED modifications is a better ground than the presence of the photon-matter axial coupling, it is also related with new physics \cite{Pospelov:2005pr}.

We could keep going with further examples, but we believe that we made our point sufficiently clear: none of the axial fermionic bilinears necessary to generate the CP-violating term $\pazocal{F}\pazocal{G}$ can be formulated from any viable modification of QED. The Lagrangian \eqref{eq:a2}, considered in the present letter and in \cite{Yamashita:2017}, and as well the Lagrangian \eqref{eq_dip} considered in \cite{Gorghetto:2021luj}, suffer by the presence of the well-known triangle-anomaly; see, e.g. \cite{Bardeen:1969md}, or any book on this subject.

 Hence, we conclude that the presented so far models of a modified QED, containing an axial fermionic bilinear in their Lagrangian that can generate the CP-violating term in the photon sector, are not theoretically acceptable extensions of QED.
This poses an important question: if the CP-violating term $\pazocal{F}\pazocal{G}$ is being phenomenologically examined in several studies, it is imperative to have a physically consistent framework where this term can be systematically obtained through quantum corrections in the same way as the CP-invariant terms in the ordinary Euler-Heisenberg action.


\section{Conclusion}
\label{conc}

In this paper, we have examined the generation of the CP-violating extension of the Euler-Heisenberg effective action in terms of minimal modifications of QED.
We started by arguing that, if we wish to perturbatively generate the CP-violating terms in the photon effective action,
it is necessarily to consider axial fermionic bilinears modifications of the QED.

We have explicitly considered two models of photon-matter couplings: i) a parity-violating QED in terms of an axial-vector bilinear and ii) a QED added by dipole operators, the electron's electric and magnetic dipole moments.
Although these couplings are sufficient to correctly generate the CP-odd term $\pazocal{F}\pazocal{G}$ they both fail in the consistency analysis of the given models.
The parity-violating model implies in the change of the value of the electromagnetic coupling, which is known with high accuracy experimental data; thus, any change in the photon-electron coupling is severely constraint.
On the other hand, the second model where dipole operators are present, fail because the electron's electric dipole moment is not present in QED at the tree level, but arise only through quantum processes (CP-violating components of the CKM matrix).

Only these drawbacks would be enough to rule out these models which are potential candidates to consistently generate the CP-violating term of the photon sector.
However, both models suffer from an even worst problem: they are anomalous theories.
It is well known that every axial modification of QED is anomalous, rendering thus inconsistencies in the formulation of the model, associated with the breaking of unitarity and renormalizability.
Thus, we can conclude that the generation of a CP-violating effective theory in the photon sector from a viable modification  of QED is indeed impossible.

\subsection*{Acknowledgements}

We would like to express our especial thanks to M. Chaichian for his valuable comments with explanation and many illuminating discussions.
Also, we appreciate the insightful comments from M.M. Sheikh-Jabbari and M. Mohammadi.
R.B. acknowledges partial support from Conselho
Nacional de Desenvolvimento Cient\'ifico e Tecnol\'ogico (CNPq Projects No. 305427/2019-9 and No. 421886/2018-8) and Funda\c{c}\~ao de
Amparo \`a Pesquisa do Estado de Minas Gerais (FAPEMIG Project No. APQ-01142-17).


\global\long\def\link#1#2{\href{http://eudml.org/#1}{#2}}
 \global\long\def\doi#1#2{\href{http://dx.doi.org/#1}{#2}}
 \global\long\def\arXiv#1#2{\href{http://arxiv.org/abs/#1}{arXiv:#1 [#2]}}
 \global\long\def\arXivOld#1{\href{http://arxiv.org/abs/#1}{arXiv:#1}}

{}

\end{document}